\documentclass[11pt]{article}

\usepackage[agsm]{harvard}
\citationmode{abbr}

\newcommand{\abs}[1]{\left\vert #1 \right\vert}
\newcommand{\norm}[1]{\left\Vert #1 \right\Vert}
\newcommand{\beqn}{\begin{eqnarray*}}
\newcommand{\eeqn}{\end{eqnarray*}}
\newcommand{\be}[1]{\begin{equation}\label{#1}}
\newcommand{\ee}{\end{equation}}
\newcommand{\halmos}{\rule{1ex}{1.4ex}}
\newcommand{\qed}{\hfill \halmos} 

\newcommand{\R}{{\mathbb R}}  
\newcommand{\N}{{\mathbb N}}  

\usepackage[dvips]{graphicx}
\usepackage[utf8]{inputenc}
\usepackage{psfrag}
\usepackage{amsthm}
\usepackage{amsmath}
\usepackage{amssymb}

\newtheorem{Example}{Example}
\newtheorem{Remark}{Remark}
\newtheorem{Theorem}{Theorem}
\newtheorem{Definition}{Definition}
\newtheorem{Lemma}{Lemma}

\title{Global entrainment of transcriptional systems to periodic inputs}
\author{Giovanni Russo\thanks{Dept. of Systems and Computer Engineering,
    University of Naples Federico II, Naples, Italy},  Mario di
  Bernardo$^{*\ddag}$, Eduardo D. Sontag$^{\dag}$\thanks{Dept. of Mathematics, Rutgers University, United States}\thanks{Correspondence: mario.dibernardo@unina.it (M.d.B.), sontag@math.rutgers.edu (E.  S.)}}

\begin{document}
  
\maketitle

\subsection*{ABSTRACT}

This paper addresses the problem of giving conditions for transcriptional systems to be globally entrained to external periodic inputs. By using contraction theory, a powerful tool from dynamical systems theory, it is shown that certain systems driven by external periodic signals have the property that all solutions converge to a fixed limit cycle.  General
results are proved, and the properties are verified in the specific case of
some models of transcriptional systems.

\section{Introduction}

Periodic, clock-like rhythms pervade nature and regulate the function of all
living organisms. For instance, \emph{circadian rhythms} are regulated by an
endogenous biological clock entrained by the light signals from the
environment that then acts as a pacemaker,
\cite{Gon_Ber_Wal_Kra_Her_05}. Moreover, such an entrainment can be obtained
even if daily variations are present, like e.g. temperature and light
variations. Another important example of entrainment in biological systems is
at the molecular level, where the synchronization of several cellular
processes is regulated by the cell cycle \cite{Tys_Csi_Now_02}. 

An important question in mathematical and computational biology is
that of finding conditions ensuring that entrainment occurs. 
The objective is to identify classes of biological systems that can be
entrained by an exogenous signal. 
To solve this problem, modelers often resort to simulations
in order to show the existence of periodic solutions in the system of
interest.  Simulations, however, can never prove that solutions will exist for
all parameter values, and they are subject to numerical errors. Moreover,
robustness of entrained solutions needs to be checked in the presence of noise
and uncertainties, which cannot be avoided experimentally.

From a mathematical viewpoint, the problem of formally showing that
entrainment takes place is known to be very difficult. Indeed, if a
stable linear time-invariant model is used to represent the system of
interest, then entrainment is usually expected, when the system is driven by
an external periodic input, with the system response being a filtered, shifted
version of the external driving signal.  However, in general, as is often the
case in biology, models are nonlinear.  
The response of nonlinear systems to periodic inputs is the subject of much
current systems biology experimentation; for example,
in \cite{MeMu:08}, the case of a cell signaling system driven by
a periodic square-wave input is considered.
From measurements of a periodic output, the authors fit a transfer function
to the system, implicitly modeling the system as linear even though 
(as stated in the Suppemental Materials to \cite{MeMu:08}) there are
saturation effects so the true system is nonlinear.
For nonlinear systems, driving the system by an external
periodic signal does not guarantee the system response to also be a periodic
solution, as nonlinear systems can exhibit harmonic generation or suppression
and complex behaviour such as chaos or quasiperiodic solutions \cite{Ku:98}.
This may happen even if the system is well-behaved with respect to
constant inputs; for example, there are systems which converge to a fixed
steady state no matter what is the input excitation, so long as this input
signal is constant, yet respond chaotically to the simplest oscillatory
input; we outline such an example in an Appendix to this paper, see
also~\cite{eds:arxiv09}. 
Thus, a most interesting open problem is that of finding conditions for the
entrainment to external inputs of biological systems modelled by sets of
nonlinear differential equations.

One approach to analyzing the convergence behavior of nonlinear dynamical
systems is to use Lyapunov functions. However, in biological applications, the
appropriate Lyapunov functions are not always easy to find and, moreover,
convergence is not guaranteed in general in the presence of noise and/or
uncertainties.  Moreover, such an approach can be hard to apply to the case of
non-autonomous systems (that is, dynamical systems directly dependent on
time), as is the case when dealing with periodically forced systems.

The above limitations can be overcome if the convergence problem is interpreted
as a property of all trajectories, asking that all solutions converge towards
one another (contraction).
This is the viewpoint of \emph{contraction theory}, \cite{Loh_Slo_98},
\cite{Loh_Slo_00}, and more generally incremental stability methods
\cite{Ang_02}.
Global results are possible, and these are robust to noise, in the sense that,
if a system satisfies a contraction property then trajectories remain bounded
in the phase space \cite{Pha_Tab_Slo_09}.  
Contraction theory has a long history.  Contractions in
metric functional spaces can be traced back to the work of Banach and
Caccioppoli \cite{Gra_03} and, in the field of dynamical systems,
to~\cite{Hartmann} and even to~\cite{Lewis} (see also~\cite{Pav_Pog_Wou_Nij},
~\cite{Ang_02}, and e.g.~\cite{pde} for a more exhaustive list of related
references).  Contraction theory has been successfully applied to both
nonlinear control and observer problems, \cite{Loh_Slo_00}, \cite{Jou_04_a}
and, more recently, to synchronization and consensus problems in complex
networks \cite{Slo_Wan_Rif_98}, \cite{Wan_Slo_05}. In \cite{Rus_diB_09b} it was proposed that contraction can be particularly useful when dealing with the analysis and characterization of biological networks. In particular, it was found that using non Euclidean norms can be particularly effective in this context \cite{Rus_diB_09b}, \cite{Rus_diB_09}.

One of the objectives of this paper is to give a self-contained exposition,
with all proofs included, of results in contraction theory as applied to
entrainment of periodic signals, and, moreover, to show their applicability to
a problem of biological interest, having to do with a driven transcriptional
system.  A surprising fact is that, for these applications, and contrary to many
engineering aplications, norms other than Euclidean, and associated matrix
measures,  must be considered.

\subsection{Mathematical tools}

We consider in this paper systems of ordinary differential equations,
generally time-dependent:
\be{eqn:gensys}
\dot x = f(t,x)
\ee
defined for $t\in [0,\infty )$ and $x\in C$, where $C$ is a subset of $\R^n$.
It will be assumed that $f(t,x)$ is differentiable on $x$, and that
$f(t,x)$, as well as the Jacobian of $f$ with respect to $x$,
denoted as $J(t,x) = \frac{\partial f}{\partial x}(t,x)$, are both continuous
in $(t,x)$. 
In applications of the theory, it is often the case that $C$ will
be a closed set, for example given by non-negativity constraints on variables
as well as linear equalities representing mass-conservation laws.
For a non-open set $C$, differentiability in $x$ means that the vector field
$f(t,\bullet )$ can be extended as a differentiable function to some open set which
includes $C$, and the continuity hypotheses with respect to $(t,x)$ hold on
this open set.

We denote by
$
\varphi(t,s,\xi )
$
the value of the solution $x(t)$ at time $t$ of the differential
equation~(\ref{eqn:gensys}) with initial value $x(s)=\xi $.
It is implicit in the notation that $\varphi(t,s,\xi )\in C$ (``forward invariance''
of the state set $C$).
This solution is in principle defined only on some interval $s\leq t<s+\varepsilon $,
but we will assume that $\varphi(t,s,\xi )$ is defined for all $t\geq s$.
Conditions which guarantee such a ``forward-completeness'' property
are often satisfied in biological applications, for example whenever
the set $C$ is closed and bounded, or whenever the vector field $f$ is
bounded.  (See Appendix C in~\cite{mct} for more discussion, as well
as~\cite{angeli-sontag-fc} for a characterization of the forward completeness
property.) 
Under the stated assumptions, the function $\varphi$ is jointly differentiable
in all its arguments (this is a standard fact on well-posedness of
differential equations, see for example Apendix C in~\cite{mct}).

We recall (see for instance~\cite{michelbook}) that, given a vector norm on Euclidean space ($\abs{\bullet}$), with its induced matrix norm $\norm{A}$, the associated \emph{matrix measure} $\mu$ is defined as the directional derivative of the matrix norm, that is,
\[
\mu(A) \,:=\;
\lim_{h \searrow 0} \frac{1}{h} \left(\norm{I+hA}-1\right).
\]
For example, if $\abs{\bullet}$ is the standard Euclidean 2-norm, then
$\mu(A)$ is the maximum eigenvalue of the symmetric part of $A$.
As we shall see, however, different norms will be useful for our applications.
Matrix measures are also known as ``\emph{logarithmic norms}'', a concept
independently introduced by Germund Dahlquist and Sergei Lozinskii in 1959,
\cite{dahlquist,lozinskii}.
The limit is known to exist, and the convergence is monotonic,
see~\cite{strom,dahlquist}.

We will say that system (\ref{eqn:gensys}) is {\em infinitesimally
  contracting} on a convex set $C \subseteq \R ^n$ if there exists some norm
in $C$, with associated matrix measure $\mu$ such that, for some constant 
$c \in\R-\left\{ 0 \right\}$,  
\begin{equation} \label{eqn:contrcond}
\mu \left(J\left(x,t\right)\right) \le - c^2, \quad \forall x \in C, \quad \forall t \ge 0.
\end{equation}

Let us discuss very informally (rigorous proofs are given later) the motivation
for this concept.  Since by assumption $f\left(t,x\right)$ is continuously
differentiable, the following exact \emph{differential} relation can be
obtained from (\ref{eqn:gensys}): 
\begin{equation} \label{eqn:genvirt}
\delta \dot x = J\left(t,x\right) \delta x,
\end{equation}
where, as before, $J=J\left(t,x\right)$ denotes the Jacobian of the vector
field $f$, as a function of $x\in C$ and $t \in \R^+$.
(The object $\delta x$ can be thought of as a ``virtual displacement''
in the language of mechanics, as in~\cite{Arn_78}, which views
such displacements as linear tangent differential forms differentiable with
respect to time.) Consider now two neighboring trajectories of (\ref{eqn:gensys}), evolving in
$C$, and the virtual displacements between them. Note that (\ref{eqn:genvirt})
can be seen as a linear time-varying dynamical system of the form:
$$
\delta \dot x = J\left(t\right)\delta x.
$$
Hence, an upper bound for the magnitude of its solutions can be obtained by
means of the Coppel inequality \cite{Vid_93}, yielding: 
\begin{equation} \label{eqn:coppelvirt}
\abs{\delta x} \le \abs{\delta x_0}  e^{\int_{0}^t \mu \left( J\left(\xi\right)\right)d\xi},
\end{equation}
where $\mu \left(J\right)$ is the matrix measure of the system Jacobian induced by the norm being considered on the states and $\abs{\delta x \left(0\right)} = \abs{\delta x_0}$.
Using (\ref{eqn:coppelvirt}) and (\ref{eqn:contrcond}), we have that
\[
\exists \quad \beta >0 :\quad \abs{\delta x\left(t\right)} \le \beta e^{-c^2 t}.
\]
Thus, trajectories starting from infinitesimally close initial conditions converge exponentially towards each other.
In what follows we will refer to $c^2$ as \emph{contraction (or convergence) rate}.

The key theoretical result about contracting systems links infinitesimal and
global contractivity, and is stated below.  This result can be traced, under
different technical assumptions, to e.g. \cite{Loh_Slo_98},
\cite{Pav_Pog_Wou_Nij}, \cite{Lewis}, \cite{Hartmann}. 

\begin{Theorem} \label{theo:main}
Suppose that $C$ is a convex subset of $\R^n$ and that $f(t,x)$ is
infinitesimally contracting with contraction rate $c^2$.
Then, for every two solutions $x(t)=\varphi(t,0,\xi )$ and $z(t)=\varphi(t,0,\zeta )$
of~(\ref{eqn:gensys}), it holds that:
\be{eqn:contract}
\abs{x(t)-z(t)} \;\leq \; e^{-c^2t} \abs{\xi -\zeta }, 
\quad\quad\forall\,t\ge 0\,.
\ee
\end{Theorem}

In other words, infinitesimal contractivity implies global contractivity.
In the Appendix, we provide a self-contained proof of Theorem~\ref{theo:main}.
In fact, the result is shown there in a generalized form, in which convexity
is replaced by a weaker constraint on the geometry of the space.

In actual applications, often one is given a system which depends implicitly on the time, $t$, by means of a continuous function $u\left(t\right)$, i.e. systems dynamics are represented by $\dot x=f\left(x,u\left(t\right)\right)$. In this case, $u\left(t\right) : \R^+ \rightarrow U$ (where $U$ is some subset of $\R$), represents an external input.
It is important to observe that the contractivity
property does not require any prior information about this external input.
In fact, since $u\left(t\right)$ does not depend on the system state variables,  when checking the property, it may be viewed as a constant parameter, $u \in U$. Thus, if contractivity of $f\left(x,u\right)$ holds uniformly $\forall u \in U$, then it will also hold for $f\left(x,u\left(t\right)\right)$.

Given a number $T>0$, we will say that system~(\ref{eqn:gensys}) is
\emph{$T$-periodic} if it holds that 
\[
f(t+T,x) \,=\, f(t,x) 
\quad\quad\forall\,t\geq 0,\,x\in C\,.
\]
Notice that the system $\dot x=f\left(x,u\left(t\right)\right)$ is $T$-periodic, if the external input, $u\left(t\right)$, is itself a periodic function of period $T$.

The following is the basic theoretical result about periodic orbits that will
be used in the paper.  It may be found, under various different technical
variants, in the references given above.

\begin{Theorem}\label{theorem:contraction}
Suppose that:
\begin{itemize}
\item $C$ is a closed convex subset of $\R^n$;
\item $f$ is infinitesimally contracting with contraction rate $c^2$;
\item $f$ is $T$-periodic.
\end{itemize}
Then, there is a unique periodic solution $\alpha (t):[0,\infty )\rightarrow C$ of~(\ref{eqn:gensys})
of period $T$ and,
for every solution $x(t)$, it holds that $\abs{x\left(t\right)-\alpha \left(t\right)}\rightarrow 0$ as $t\rightarrow \infty $.
\end{Theorem}

In the Appendix of this paper, we provide a self-contained proof of
Theorem~\ref{theorem:contraction}, in a generalized form which does not
require convexity.

\subsection{A simple example}

As a first example to illustrate the application of the concepts introduced so
far, we choose a simple bimolecular reaction, in which a molecule of $A$ and
one of $B$ can reversibly combine to produce a molecule of $C$.

This system can be modeled by the following set of differential equations:
\begin{equation} \label{eqn:t-cell_example}
\begin{array}{*{20}l}
\dot A = -k_1 AB + k_{-1}C, \\
\dot B = -k_1 AB + k_{-1}C, \\
\dot C = k_1 AB - k_{-1} C, \\
\end{array}
\end{equation}
where we are using $A=A(t)$ to denote the concentration of $A$ and so forth.
The system evolves in the positive orthant of $\R^3$.
Solutions satisfy (stoichiometry) constraints:
\begin{equation}\label{eqn:sto_class_example}
\begin{array}{*{20}c}
A(t)+C(t) = \alpha \\
B(t)+C(t) = \beta  \\
\end{array}
\end{equation}
for some constants $\alpha$ and $\beta$.

We will assume that one or both of the ``kinetic constants'' $k_i$ are
time-varying, with period $T$.  Such a situation arises when the $k_i$'s
depend on concentrations of additional enzymes, which are available in
large amounts compared to the concentrations of $A,B,C$, but whose
concentrations are periodically varying.
The only assumption will be that $k_1(t)\geq k_1^0>0$ and
$k_{-1}(t)\geq k_{-1}^0>0$ for all $t$.

Because of the conservation laws~(\ref{eqn:sto_class_example}), we may
restrict our study to the equation for $C$.  Once that all solutions
of this equation are shown to globally converge to a periodic orbit,
the same will follow for $A(t)=\alpha-C(t)$ and $B(t)=\beta-C(t)$.
We have that:
\begin{equation}
\dot C = k_1 \left(\alpha - C\right)\left(\beta - C\right) - k_{-1}C.
\end{equation}
Because $A(t)\geq0$ and $B(t)\geq0$, this system is studied on the subset
of $\R$ defined by $0\leq C \leq \mbox{min}\left\{\alpha,\beta\right\}$.
The equation can be rewritten as:
\begin{equation} \label{eqn:t-cell_example_res}
\dot C = k_1 \left(\alpha \beta - \alpha C- \beta C + C^2\right) - k_{-1}C.
\end{equation}
Differentiation with respect to $C$ of the right-hand side in the above system
yields this ($1 \times 1$) Jacobian: 
\begin{equation} \label{eqn:t-cell_example_J}
J:=k_1 \left( -\left(\alpha +\beta \right)+ 2 C -k_{-1}\right).
\end{equation}
Since we know that $-\alpha+C\leq0$ and $-\beta+C\leq0$, it follows that
\[
J \leq -k_1 k_{-1} \leq -k_1^0 k_{-1}^0 := -c^2
\]
for $c = \sqrt{k_1^0 k_{-1}^0}$.  Using any norm (this example is in dimension
one) we have that $\mu(J)<-c^2$.
So~(\ref{eqn:t-cell_example}) is contracting and, by means of Theorem
\ref{theorem:contraction}, solutions will globally converge to a unique solution
of period $T$ (notice that such a solution depends on system parameters).

\begin{figure}[thbp]
\begin{center}
\centering \psfrag{x}[c]{{time (minutes)}}
\centering \psfrag{y}[c]{{$C$ (arbitrary units)}}
  \includegraphics[width=8cm]{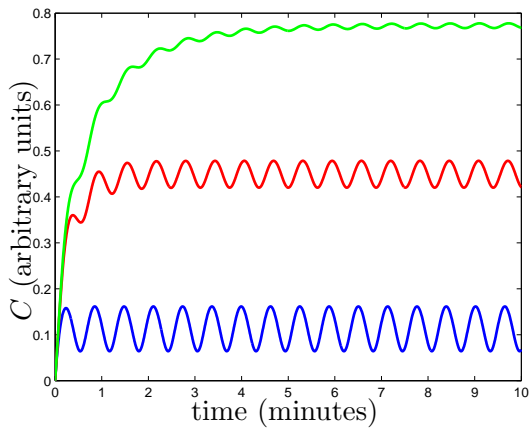}
  \caption{Entrainment of (\ref{eqn:t-cell_example_res}) to the periodic input $u(t)=1.5+\sin (10 t)$ for $k_{-1} = 10$ (blue), $k_{-1} = 1$ (green), $k_{-1} = 0.1$ (red). Notice that an increase of $k_{-1}$, causes an increase of the contraction rate, hence trajectories converge faster to the system unique periodic attractor. The other system parameters are set to: $\alpha = \beta =1$, $k_2 = 0.1$.}
  \label{fig:simple_bio_sim}
  \end{center}
\end{figure}

Figure \ref{fig:simple_bio_sim} shows the behavior of the dynamical system (\ref{eqn:t-cell_example_res}), using two different values of $k_{-1}$. Notice that the asymptotic behavior of the system depends on the particular choice of the biochemical parameters being used. Furthermore, it is worth noticing here that the higher the value of $k_{-1}$, the faster will be the convergence to the attractor.

\section{Results}

\subsection{Mathematical model and problem statement}

We study a general externally-driven transcriptional module.
We assume that the rate of production of a transcription factor $X$ is
proportional to the value of a time dependent input function $u(t)$, and $X$
is subject to degradation and/or dilution at a linear rate.  (Later, we
generalize the model to also allow nonlinear degradation as well.)
The signal $u(t)$ might be an external input, or it might represent the
concentration of an enzyme or of a second messenger that activates $X$.
In turn, $X$ drives a downstream transcriptional module by binding
to a promoter (or substrate), denoted by $E$, whose free concentration is
denoted as $e=e(t)$. The binding reaction of $X$ with
$E$ is reversible and given by: 
\[
X+E \rightleftharpoons Y,
\]
where $Y$ is the complex protein-promoter, and the binding and dissociation
rates are $k_1$ and $k_2$ respectively.  As the promoter is not
subject to decay, its total concentration, $E_{T}$, is conserved, so
that the following conservation relation holds:
\begin{equation} \label{balance}
E+Y=E_T.
\end{equation}
We wish to study the behavior of solutions of the system that couples $X$ and
$E$, and specifically to show that, when the input $u(t)$ is periodic with
period $T$, this coupled system has the property that all solutions converge
to some globally attracting limit cycle whose period is also $T$.

\begin{figure}[htl]
\centering
  \includegraphics[scale=.5]{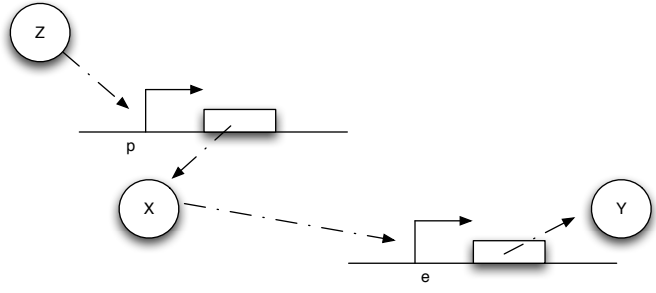}
  \caption{A schematic diagram of the two transcriptional modules modeled in (\ref{model})}
  \label{transcriptional}
\end{figure}

Such transcriptional modules are ubiquitous in biology, natural as well as
synthetic, and their behavior was recently studied in~\cite{DelV_Nin_Son_08}
in the context of ``retroactivity'' (impedance or load) effects.
If we think of $u(t)$ as the concentration of  a protein $Z$ that is a
transcription factor for $X$, and we ignore fast mRNA dynamics, such a system
can be schematically represented as in Figure \ref{transcriptional},
which is adapted from~\cite{DelV_Nin_Son_08}.
Notice that $u(t)$ here does not need to be the concentration of a
transcriptional activator of $X$ for our 
results to hold.  The results will be valid for any mathematical model for the concentrations, $x$, of $X$ and $y$, of $Y$ (the
concentration of $E$ is conserved) of the form:
 \begin{equation}\label{model}
\begin{array}{*{20}l}
\dot x = u \left(t\right)-\delta x +k_1 y -  k_2 \left(E_T-y\right)x\\
\dot y= -k_1y + k_2 \left(E_T-y\right)x \,.\\
\end{array}
\end{equation}

\begin{figure}[thbp]
\begin{center}
\centering \psfrag{t}[c]{{time (minutes)}}
  \includegraphics[width=8cm]{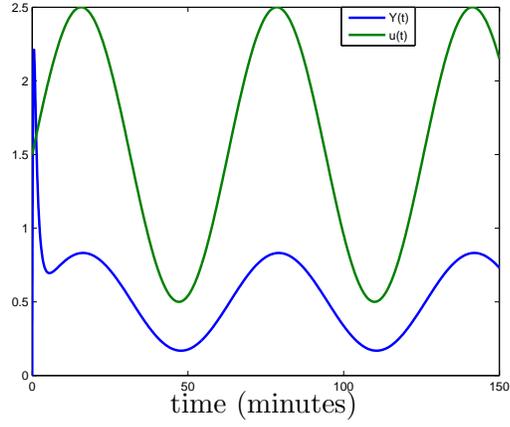}
\centering \psfrag{t}[c]{{time (minutes)}}
\psfrag{y}[c]{{Y (arbitrary units)}}
  \includegraphics[width=8cm]{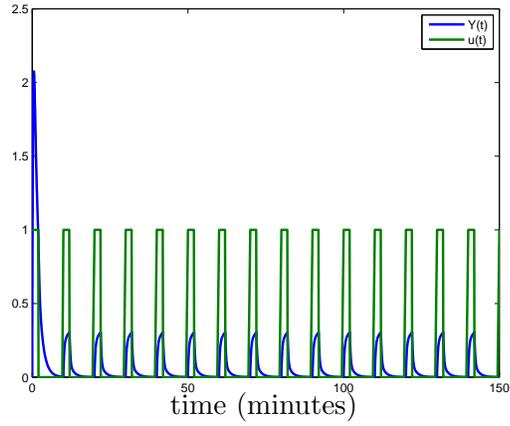}
  \caption{Entrainment of the transcriptional module (\ref{model}) output (green), $Y$, to the periodic input (blue): $u(t) = 1.5 + \sin(0.1t)$ (left) and to a repeating $\left\{0,1\right\}$ sequence (right). System parameters are set to: $\delta=3$, $k_1$=1, $k_2=0.1$.}
  \label{fig:modulesim}
  \end{center}
\end{figure}

Our main objective in this paper is, thus, to show that, when $u$ is a periodic
input, all solutions of system~(\ref{model}) converge to a (unique) limit
cycle (Figure \ref{fig:modulesim}).  
The key tool in this analysis is to show that, when no input is
present, the system is infinitesimally, and hence globally, contracting.

Thus, the main step will be to establish the following technical result,
see Section~\ref{sec:mainproof}:

\begin{Theorem}\label{theorem:contract}
The system
\beqn
\dot x &=& -\delta x +k_1 y -  k_2 \left(E_T-y\right)x\\
\dot y &=& -k_1y + k_2 \left(E_T-y\right)x
\eeqn
where
\begin{equation} \label{eqn:contr_region}
(x(t),y(t))\in C = [0,\infty)\times[0,E_T]
\end{equation}
for all $t\geq0$, and $E_T$, $k_1$, $k_2$, and $\delta$ are arbitrary positive
constants, is contracting.
\end{Theorem}

By means of Theorem \ref{theorem:contraction}, we then have the following immediate Corollary:

\begin{Theorem}\label{theorem:periodic}
For any given nonnegative
periodic input $u$ of period $T$, all solutions of system (\ref{model}) converge
exponentially to a periodic solution of period $T$. 
\end{Theorem}

In the following sections, we introduce a matrix measure that will help
establish contractivity, and we prove Theorem~\ref{theorem:contract}.
We will also discuss several extensions of this result, allowing the
consideration of multiple driven subsystems as well as more general
nonlinear systems with a similar structure.

\subsection{Proof of Theorem~\protect{\ref{theorem:contract}}}
\label{sec:mainproof}

We will use Theorem~\ref{theorem:contraction}.
The Jacobian matrix to be studied is:
\begin{equation}\label{eq_2}
J:=\left[\begin{array}{*{20}c}
-\delta -k_2 \left( E_T-y\right) & k_1+k_2x\\
k_2 \left( E_T-y\right) &-k_1-k_2x\\
\end{array}\right].
\end{equation}
As matrix measure, we will use the measure $\mu_{P,1}$ induced by the vector
norm $\abs{Px}_1$, where $P$ is a suitable nonsingular matrix. 
More specifically, we will pick $P$ diagonal:
\begin{equation} \label{eq_3}
\left[ \begin{array}{*{20}c}
p_1 & 0\\
0 & p_2\\
\end{array}\right],
\end{equation}
where $p_1$ and $p_2$ are two positive numbers to be appropriately chosen
depending on the parameters defining the system.

It follows from general facts about matrix norms that
\begin{equation} \label{eq_4}
\mu_{P,1}\left(J\right)=\mu_1\left(PJP^{-1}\right),
\end{equation}
where $\mu_1$ is the measure associated to the $\abs{\bullet}_1$ norm and is
explicitly given by the following formula:
\begin{equation} \label{eq_5}
\mu_{1} \left(J\right)=\max_j\left( J_{jj}+ \sum_{i\ne j} \abs{J_{ij}} \right).
\end{equation}
Observe that, if the entries of $J$ are negative, then asking that
$\mu_1(J)<0$ amounts to a column diagonal dominance condition.
(The above formula is for real matrices.  If complex matrices would be
considered, then the term $J_{jj}$ should be replaced by
its real part $\Re\{J_{jj}\}$.) 

Thus, the first step in computing $\mu_{P,1}\left(J\right)$ is to
calculate 
$PJP^{-1}$:
\begin{equation}\label{eq_6}
\left[ \begin{array}{*{20}c}
-\delta -k_2 \left( E_T-y\right) & \frac{p_1}{p_2}\left(k_1+k_2x\right)\\
\frac{p_2}{p_1}\left[k_2 \left( E_T-y\right)\right] &-k_1-k_2x\\
\end{array}\right].
\end{equation}
Using (\ref{eq_5}), we obtain:
\begin{equation}\label{eq_7}
\mu_{P,1} \left(J\right)=\max 
\left\{-\delta -k_2 \left(E_T -y \right)+  \abs{ \frac{p_2}{p_1}k_2\left(E_T-y\right)} ; -k_1-k_2x + \abs{ \frac{p_1}{p_2}\left( k_1+k_2x\right)} \right\}\,.
\end{equation}
Note that we are not interested in calculating the exact value for the above
measure, but just in ensuring that it is negative. 
To guarantee that $\mu_{P,1}\left(J\right)<0$, the following two
conditions must hold:
\begin{equation}\label{eq_8}
-\delta -k_2 \left(E_T -y \right)+  \abs{ \frac{p_2}{p_1}k_2\left(E_T-y\right)}< -c_1^2\,;
\end{equation}
\begin{equation}\label{eq_9}
-k_1-k_2x + \abs{ \frac{p_1}{p_2}\left( k_1+k_2x\right)}< -c_2^2\,.
\end{equation}
Thus, the problem becomes that of checking if there exists an appropriate
range of values for $p_1$, $p_2$ that satisfy (\ref{eq_8}) and (\ref{eq_9})
simultaneously.

The left hand side of (\ref{eq_9}) can be written as:
\begin{equation}\label{eq_10}
\left(\frac{p_1}{p_2}-1\right)\left(k_1+k_2x\right),
\end{equation}
which is negative if and only if $p_1 < p_2$. In particular, in this case we have:
$$
\left(\frac{p_1}{p_2}-1\right)\left(k_1+k_2x\right) \le \left(\frac{p_1}{p_2}-1\right) k_1:= -c_1^2.
$$
The idea is now to ensure 
negativity of (\ref{eq_8}) by using appropriate values for $p_1$ and $p_2$
which fulfill the above constraint. 
Recall that the term $E_T-y\geq0$ because of the choice of the state space
(this quantity represents a concentration). 
Thus, the left hand side of (\ref{eq_8}) becomes
\begin{equation}\label{eq_11}
-\delta + \left(\frac{p_2}{p_1} -1\right)k_2 \left(E_T -y \right)
\end{equation}
The next step is to choose appropriately $p_2$ and $p_1$ (without violating
the constraint $p_2>p_1$). Imposing $p_2/p_1 = 1+\varepsilon$,
$\varepsilon>0$, (\ref{eq_11}) becomes
\begin{equation}\label{eq_11b}
-\delta +\varepsilon k_2\left(E_T-y\right).
\end{equation}
Then, we have to choose an appropriate value for $\varepsilon$ in order to make
the above quantity uniformly negative. In particular, (\ref{eq_11b}) is
uniformly negative if and only if
\begin{equation}\label{eq_12}
\varepsilon< \frac{\delta}{k_2\left(E_T-y\right)}\le \frac{\delta}{k_2E_T}.
\end{equation}
We can now choose 
$$
\varepsilon =  \frac{\delta}{k_2E_T}- \xi,
$$
with $0<\xi<\frac{\delta}{k_2E_T}$. In this case, (\ref{eq_11b}) becomes
$$
-\delta +\varepsilon k_2\left(E_T-y\right) \le -\xi k_2 E_T :=-c_2^2.
$$
Thus, choosing $p_1=1$ and $p_2=1+\varepsilon = 1+ \frac{\delta}{k_2E_T}- \xi$, with $0<\xi<\frac{\delta}{k_2E_T}$, we have
$\mu_{1,P}\left(J\right)<-c^2$. Furthermore, the contraction rate $c^2$,  is given by:
$$
\min \left\{ c_1^2, c_2^2 \right\} .
$$
Notice that $c^2$ depends on both system parameters and on the elements $p_1$, $p_2$, i.e. it depends on the particular metric chosen to prove contraction.
This completes the proof of the Theorem.
\qed

\subsection{Generalizations} \label{gen}

In this Section, we discuss various generalizations that use the same proof
technique.

\subsubsection{Assuming $X$ activation by enzyme kinetics}

The previous model assumed that $X$ was created in proportion to the amount of
external signal $u(t)$.  While this may be a natural assumption if $u(t)$
is a transcription factor that controls the expression of $X$, a different
model applies if, instead, the ``active'' form $X$ is obtained from an
``inactive'' form $X_0$, for example through a phosphorylation reaction
which is catalyzed by a kinase whose abundance is represented by $u(t)$.
Suppose that $X$ can also be constitutively deactivated.
Thus, the complete system of reactions consists of
\[
X+E \rightleftharpoons Y,
\]
together with
\[
X_0 \rightleftharpoons X
\]
where the forward reaction depends on $u$.  Since the concentrations
of $X_0+X+Y$ must remain constant, let us say at a value $X_{\mbox{tot}}$, we
eliminate $X_0$ and have:
\begin{equation} \label{eqn:transc_enzy}
\begin{array}{*{20}l}
\dot x &=& u(t)(X_{tot}-x-y)-\delta x +k_1 y -  k_2 \left(E_T-y\right)x,\\
\dot y &=& -k_1y + k_2 \left(E_T-y\right)x.
\end{array}
\end{equation}

We will prove that if $u\left(t\right)$ is periodic and positive, i.e. $u\left(t\right) \ge u_0 >0$, then a globally
attracting limit cycle exists. Namely, it will be shown, after having
performed a 
linear
coordinate transformation,  that there exists a negative matrix measure for the system of interest.

Consider, indeed, the following change of the state variables:
\begin{equation} \label{eqn:coord_transf}
x_t = x + y.
\end{equation}
The systems dynamics, then become:
\begin{equation} \label{eqn:transf_sys_dyn}
\begin{array}{*{20}l}
\dot x_t = u\left(t\right)\left(X_{tot}- x_t\right) - \delta x_t + \delta y \\
\dot y = -k_1y + k_2\left(E_T-y\right)\left(x_t-y\right)\\
\end{array}.
\end{equation}
As matrix measure, we will now use the measure $\mu_\infty$ induced by the vector norm $\abs{\bullet}_\infty$. (Notice that this time, the matrix $P$ is the identity matrix).

Given a real matrix $J$, the matrix measure $\mu_\infty \left(J\right)$ is
explicitly given by the following formula (see e.g.~\cite{michelbook}):
\begin{equation} \label{eqn:infty_measure}
\mu_\infty \left(J\right) = \max_i \left(J_{ii}+ \sum_{j \ne i}\abs{J_{ij}}\right).
\end{equation}
(Observe that this is a row-dominance condition, in contrast to the dual
column-dominance condition used for $\mu_1$.) 

Differentiation of (\ref{eqn:transf_sys_dyn}) yields the Jacobian matrix:
$$
J:=\left[ \begin{array}{*{20}c}
- u\left(t\right)-\delta & \delta \\
k_2\left(E_T-y\right) & -k_1 + k_2 \left(-E_T-x_t+2y\right)\\
\end{array} \right].
$$
Thus, it immediately follow from (\ref{eqn:infty_measure}) that $\mu_\infty \left(J\right)$ is negative if and only if:
\begin{equation}
- u\left(t\right)-\delta + \abs{\delta} <-c_1^2;
\end{equation}
\begin{equation}
-k_1 + k_2 \left(-E_T-x_t+2y\right) + \abs{k_2\left(E_T-y\right)}<-c_2^2.
\end{equation}
The first inequality is clearly satisfied since by hypotheses both system parameters and the periodic input $u\left(t\right)$ are positive. In particular, we have:
$$
- u\left(t\right)-\delta + \abs{\delta} <-u_0:=-c_1^2;
$$

By using (\ref{eqn:coord_transf}) (recall that $E_T - y \ge 0$), the right hand side of the second inequality can be written as:
$$
-k_1 + k_2 \left(-E_T-x_t+2y\right) + k_2\left(E_T-y\right) = -k_1 -k_2 x.
$$
Since all system parameters are positive and $x \ge 0$, the above quantity is negative and upper bounded by $-k_1 :=-c_2^2$.

Thus, we have that $\mu_\infty \left(J\right)<-c^2$, where:
$$
c^2 = \min \left\{c_1^2,c_2^2\right\} .
$$
The contraction property for the system is then proved. By means of Theorem \ref{theorem:contraction}, we can then conclude that the system can be entrained by any periodic input.

\begin{figure}[thbp]
\begin{center}
\centering \psfrag{x}[c]{{time (minutes)}}
\centering \psfrag{y}[c]{{Y (arbitrary units)}}
  \includegraphics[width=8cm]{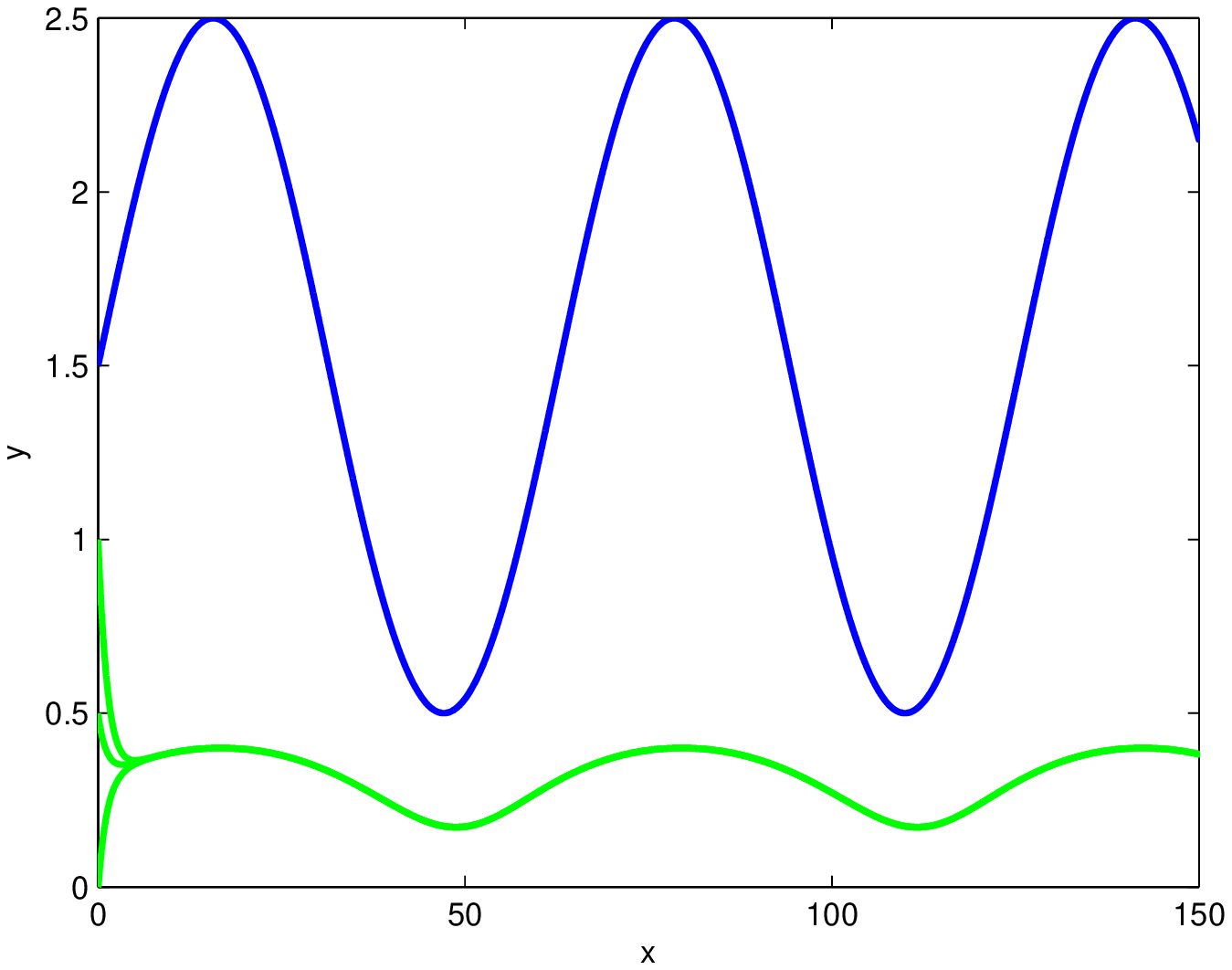}
  \centering \psfrag{x}[c]{{time (minutes)}}
\psfrag{y}[c]{{Y (arbitrary units)}}
  \includegraphics[width=8cm]{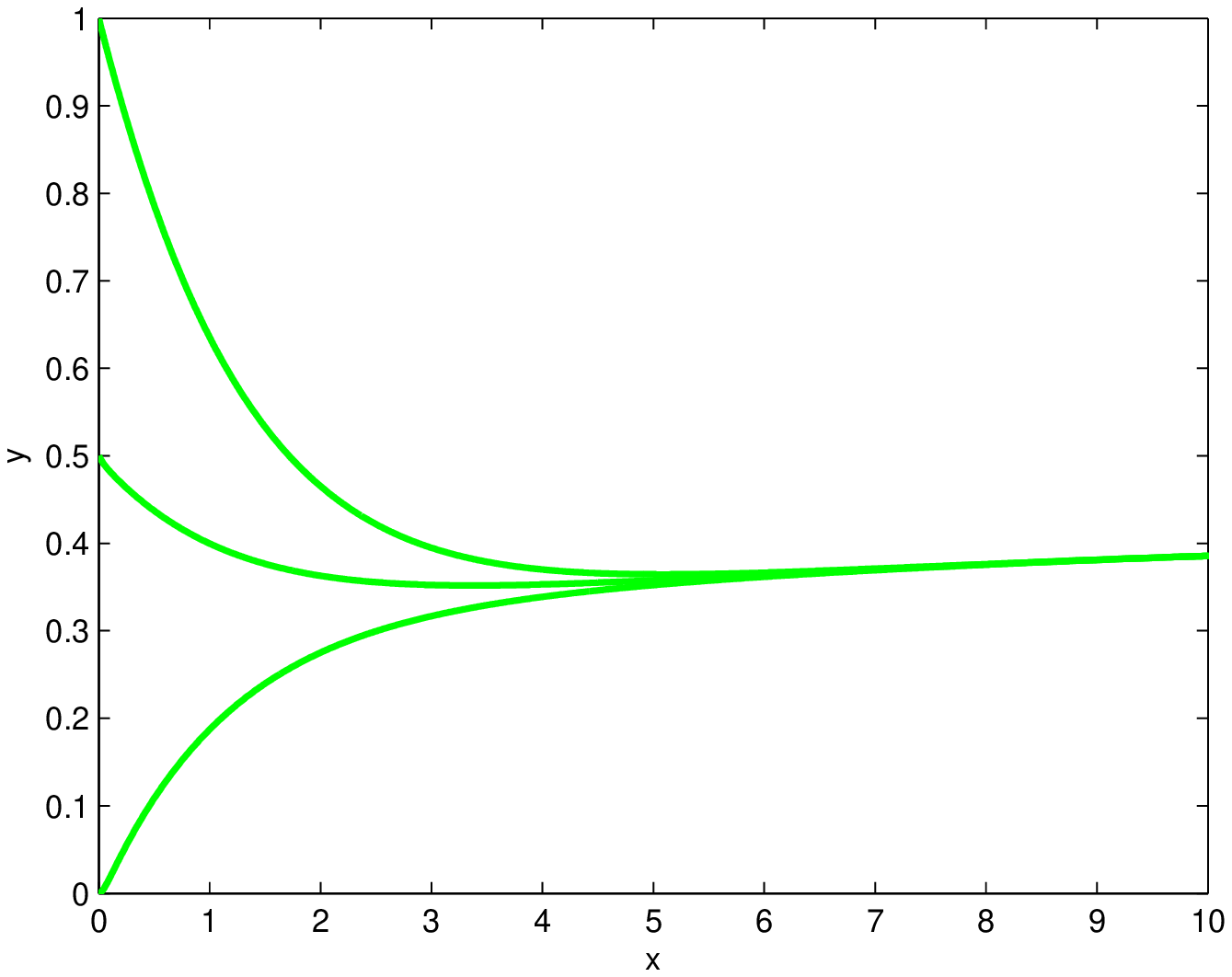}
  \caption{Left panel: entrainment of the transcriptional module (\ref{eqn:transc_enzy}) output (green), $Y$, to the periodic input (blue): $u(t) = 1.5 + \sin(0.1t)$. Right panel: zoom on $t \in \left[0,10\right]$ min. showing trajectories starting from different initial conditions converging towards the attracting limit cycle. System parameters are set to: $k_1=0.5$, $k_2=5$, $X_{tot}=1$, $E_T=1$, $\delta = 20$.}
  \label{fig:modulesim_enz}
  \end{center}
\end{figure}

Simulation results are presented in Figure \ref{fig:modulesim_enz}, where the presence of a stable limit cycle having the same period as $u\left(t\right)$ is shown.

\subsubsection{Multiple driven systems}

\begin{figure}[thpb]
\centering
  \includegraphics[scale=.5]{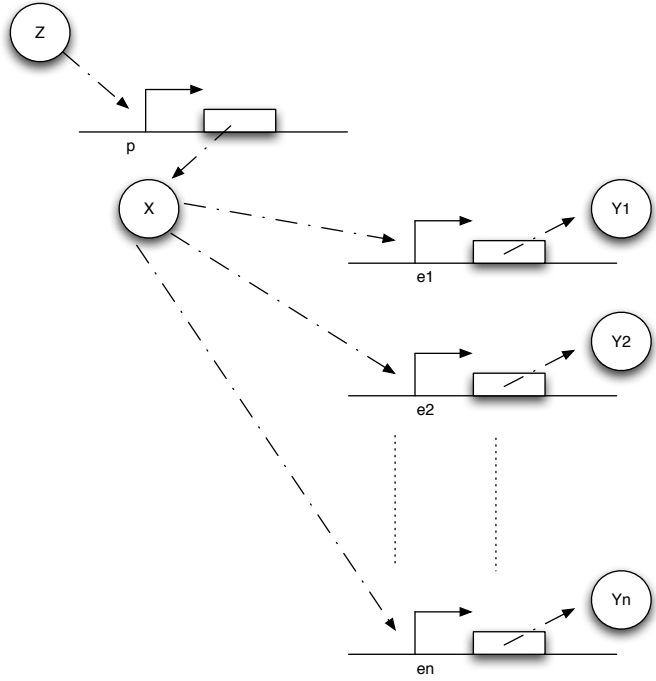}
  \caption{A schematic diagram of the transcriptional modules given in (\ref{model})}
  \label{transcriptional_2}
\end{figure}

We may also treat the case in which the species $X$ regulates multiple
downstream transcriptional modules which act independently from each other, as
shown in Figure~\ref{transcriptional_2}.
The biochemical parameters defining the different downstream modules may be
different from each other, representing a situation in which the transcription
factor $X$ regulates different species.
After proving a general result on oscillations, and assuming that parameters
satisfy the retroactivity estimates discussed in~\cite{DelV_Nin_Son_08}, one
may in this fashion design a single input-multi output module in which
e.g. the outputs are periodic functions with different mean values, settling
times, and so forth.

We denote by $E_1,\ldots,E_n$ the various promoters, and use $y_1,\ldots,y_n$
to denote the concentrations of the respective promoters complexed with $X$.
The resulting mathematical model becomes:
\begin{equation} \label{extended_model}
\begin{array}{*{20}l}
\dot x \;=\; u(t)-\delta x + K_{11}y_1-K_{21}(E_{T,1}-y_1)x \,+ \\ 
\quad\quad\quad
+\,K_{12}y_2-K_{22}(E_{T,2}-y_2)x \,+\, \cdots\\
\quad\quad\quad+\,   K_{1n}y_n-K_{2n}(E_{T,n}-y_n)x\\
\dot y_1= - K_{11}y_1+K_{21}(E_{T,1}-y_1)x\\
 \vdots \\
 \dot y_n =- K_{1n}y_n+K_{2n}(E_{T,n}-y_n)x \,.\\
\end{array}
\end{equation}

We consider the corresponding system with no input first, assuming that the
states satisfy $x(t)\geq0$ and $0\leq y_i(t)\leq E_{T,i}$ for all $t,i$.

Our generalization can be stated as follows:
\begin{Theorem} \label{extended_model_thm}
System (\ref{extended_model}) with no input (i.e. $u(t) = 0$) is contracting. Hence, if $u(t)$ is a non-zero periodic input, its solutions exponentially converge towards a periodic orbit of the same period as $u(t)$.
\end{Theorem}

\proof
We only outline the proof, since it is similar to the proof of
Theorem~\ref{theorem:periodic}.
We employ the following matrix measure:
\begin{equation} \label{eq_extended_proof_1}
\mu_{P,1}\left(J\right)=\mu_1\left(PJP^{-1}\right),
\end{equation}\label{eq_extended_proof_2}
where
\begin{equation}
P:=\left[ \begin{array}{*{20}c}
p_1 & 0 & 0 & \ldots & 0 \\
0 & p_2 & 0 & \ldots & 0 \\
\vdots & \vdots & \vdots & \vdots & \vdots \\
0 & 0 & 0 & \ldots & p_{n+1}\\
\end{array} \right]
\end{equation}
and the scalars $p_i$ have to be chosen appropriately ($p_i>0, \quad \forall i=1,\ldots,n+1$).

In this case,
\begin{equation}
J:=\left[ \begin{array}{*{20}c}
-\delta-\sum_{i=1}^{n}{K_{2i}(E_{T,i}-y_{i})} & K_{11}+K_{21}x & K_{12}+K_{22}x & \ldots & K_{1n}+K_{2n}x \\
K_{21}(E_{T,1}-y_{1}) & -K_{11}-K_{21}x & 0 & \ldots & 0 \\
K_{22}(E_{T,2}-y_{2}) & 0 & -K_{12}-K_{22}x & \ldots & 0 \\
\vdots & \vdots & \vdots & \ddots & \vdots \\
K_{2n}(E_{T,n}-y_{n}) & 0 & 0 & \ldots & -K_{1n}-K_{2n}x\\
\end{array} \right]
\end{equation}
and

\begin{equation}
PJP^{-1}:=\left[ \begin{array}{*{20}c}
-\delta-\sum_{i=1}^{n}{K_{2i}(E_{T,i}-y_{i})} &\frac{p_1}{p_2}(K_{11}+K_{21}x) & \frac{p_1}{p_3}(K_{12}+K_{22}x) & \ldots & \frac{p_1}{p_{n+1}}(K_{1n}+K_{2n}x) \\
\frac{p_2}{p_1}K_{21}(E_{T,1}-y_{1}) & -K_{11}-K_{21}x & 0 & \ldots & 0 \\
\frac{p_3}{p_1}K_{22}(E_{T,2}-y_{2}) & 0 & -K_{12}-K_{22}x & \ldots & 0 \\
\vdots & \vdots & \vdots & \ddots & \vdots \\
\frac{p_{n+1}}{p_1}K_{2n}(E_{T,n}-y_{n}) & 0 & 0 & \ldots & -K_{1n}-K_{2n}x\\
\end{array} \right]
\end{equation}

Hence, the $n+1$ inequalities to be satisfied are:
\begin{equation}
\label{eq:NN}
-\delta-\sum_{i=1}^{n}{K_{2i}(E_{T,i}-y_{i})}+\frac{1}{p_1}\sum_{i=1}^{n}p_{i+1}\left\vert K_{2i}(E_{T,i}-y_{i})\right\vert < -c_1^2
\end{equation}
and
\begin{equation}
\label{eq:KK}
-K_{1i}-K_{2i}x+\left \vert \frac{p_1}{p_{i+1}}(K_{1i}+K_{2i})x \right \vert < -c_{i+1}^2, \qquad i=1,2,\ldots,n.
\end{equation}

Clearly, the set of inequalities above admits a solution. Indeed, the left hand side of (\ref{eq:KK}) can be recast as
$$
\left(\frac{p_1}{p_{i+1}}-1\right)(K_{1i}+K_{2i}x), \qquad i=1,2,\ldots,n
$$
which is negative definite if and only if $p_1 / p_{i+1} < 1$ for all $i=1,\ldots,n$. Specifically, in this case we have
$$
\left(\frac{p_1}{p_{i+1}}-1\right)(K_{1i}+K_{2i}x) \le \left(\frac{p_1}{p_{i+1}}-1\right)K_{1i}:=-c_{i+1}^2, \qquad i=1,2,\ldots,n
$$
Also, from (\ref{eq:NN}), as $E_{T,i}-y_i \geq 0$ for all $i$, we have that (\ref{eq:NN}) can be rewritten as:
$$
-\delta-\sum_{i=1}^{n}K_{2i}(E_{T,i}-y_i)+ \sum_{i=1}^{n} \frac{p_{i+1}}{p_1} (E_{T,i}-y_i)<-c_1^2.
$$
Since $p_1 / p_{i+1} < 1$, we can impose $p_{i+1}/p_1 =1+ \varepsilon_{1,i+1}$ (with $\varepsilon_{1,i+1}>0$) and the above inequality becomes
$$
-\delta + \sum_{i=1}^{n}\varepsilon_{1,i+1}K_{2i}(E_{T,i}-y_i)<-c_1^2.
$$
Clearly, such inequality is satisfied if we choose $\varepsilon_{1,i+1}$ sufficiently small; namely:
$$
\varepsilon_{1, i+1} < \frac{\delta}{\left(n-1\right)k_2 E_{T,i}}.
$$

Following a similar derivation to that of Section \ref{sec:mainproof}, we can choose
$$
\varepsilon_{i+1} = \frac{\delta}{\left(n-1\right)k_2 E_{T,i}} - \xi _{i+1},
$$
with $0< \xi_{i+1} < \frac{\delta}{\left(n-1\right)k_2 E_{T,i}}$. In this case, we have:
$$
c_1^2 := - \sum_{i=1}^n \frac{\xi_{i+1}}{n-1}K_{2i}E_{Ti}.
$$
Thus, $\mu\left(J\right)<-c^2$, where
$$
c^2 = \min _i \left\{c_i\right\}, \quad i=1, \ldots, n+1.
$$
The second part of the theorem is then proven by applying Theorem \ref{theorem:contraction}. \endproof

\begin{figure}[thbp]

\begin{center}
\centering \psfrag{x}[c]{{time (minutes)}}
\centering \psfrag{y}[c]{{$Y_1$ (arbitrary units)}}
  \includegraphics[width=8cm]{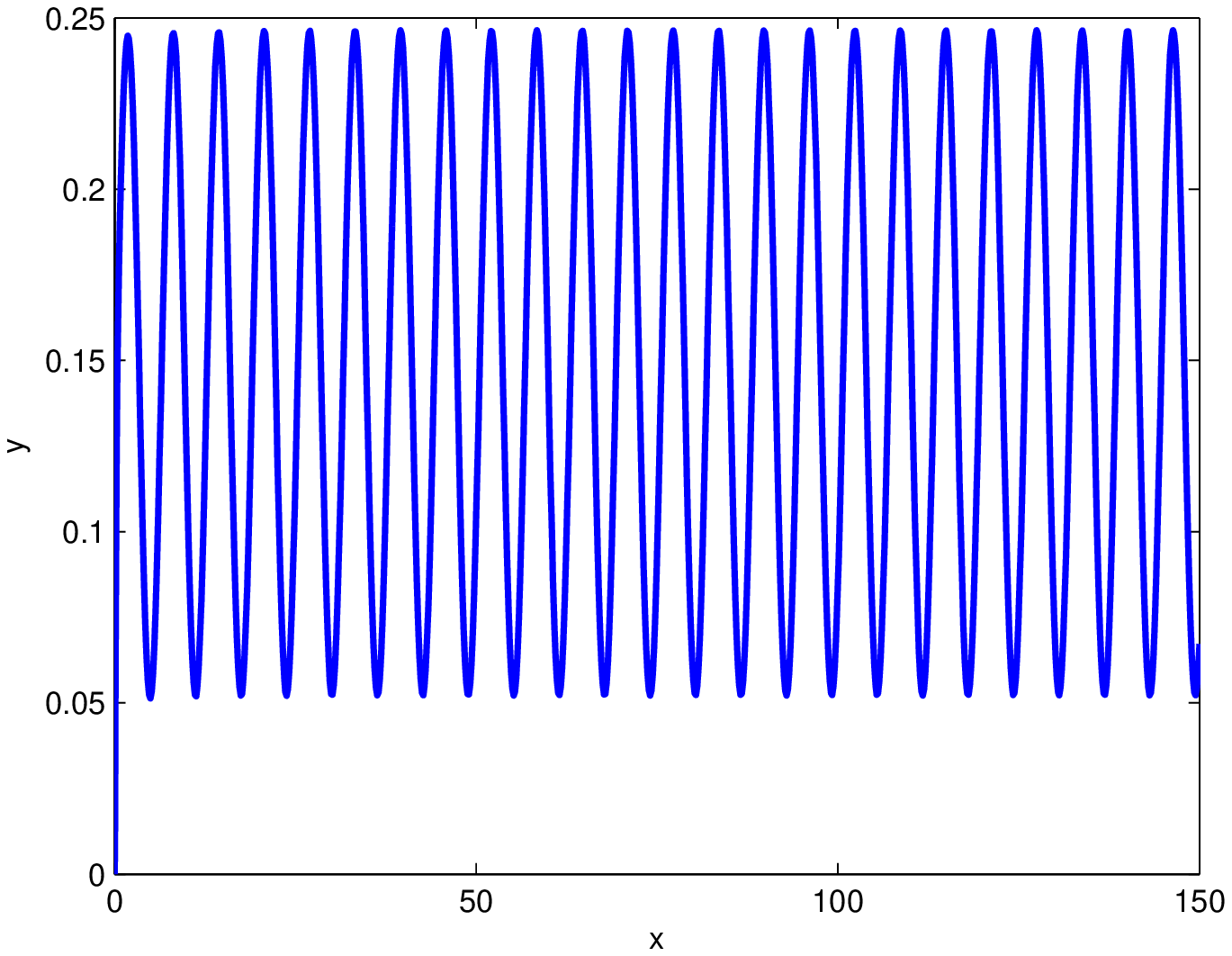}
  \centering \psfrag{x}[c]{{time (minutes)}}
\psfrag{y}[c]{{$Y_2$ (arbitrary units)}}
  \includegraphics[width=8cm]{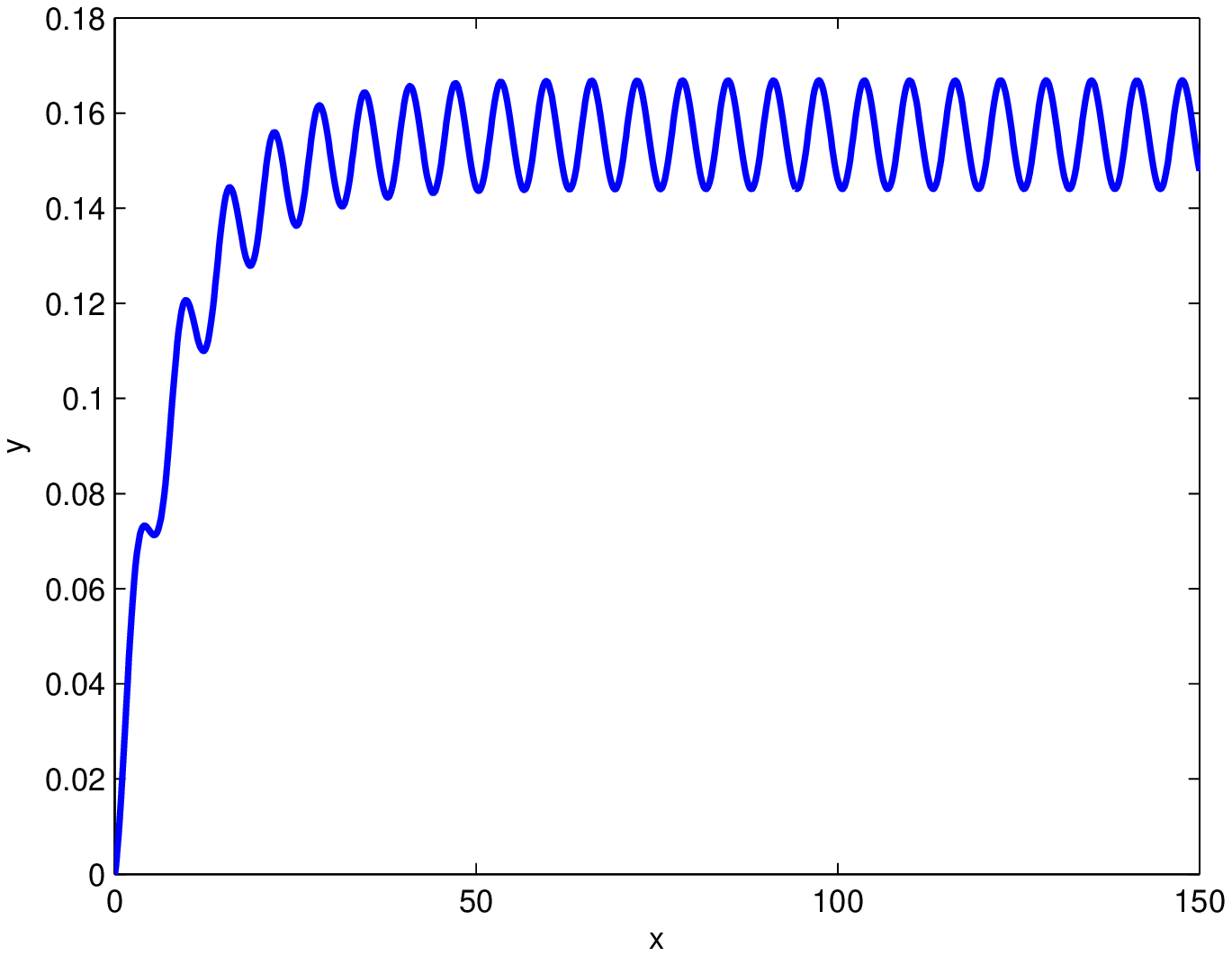}
  \caption{Outputs $Y_1$ and $Y_2$ of two transcriptional modules driven by the external periodic input $u(t)=1.5+\sin (t)$. The parameters are set to: $\delta =0.01$, $k_{11}=10$, $k_{21}=10$, $E_{T,1}=1$ for module $1$ and $k_{12}=0.1$, $k_{22} = 0.1$, $E_{T,2}=1$ for module $2$.}
  \label{fig:multiple_driven}
  \end{center}
\end{figure}

In Figure \ref{fig:multiple_driven} the behavior of two-driven downstream
transcriptional modules is shown. Notice that both the downstream modules are
entrained by the periodic input $u\left(t\right)$, but their steady state
behavior is different.

Notice that, by the same arguments used above, it can be proven that
\begin{equation} \label{extended_model_"}
\begin{array}{*{20}l}
\dot x \;=\; u(t)\left(X_{TOT}-x-\sum_{i=1}^n y_i\right)-\delta x + K_{11}y_1-K_{21}(E_{T,1}-y_1)x \,+ \\ 
\quad\quad\quad
+\,K_{12}y_2-K_{22}(E_{T,2}-y_2)x \,+\, \cdots\\
\quad\quad\quad+\,   K_{1n}y_n-K_{2n}(E_{T,n}-y_n)x\\
\dot y_1= - K_{11}y_1+K_{21}(E_{T,1}-y_1)x\\
 \vdots \\
 \dot y_n =- K_{1n}y_n+K_{2n}(E_{T,n}-y_n)x \,.\\
\end{array},
\end{equation}
is contracting.

\subsubsection{Transcriptional cascades}

\begin{figure}[thpb]
\centering
  \includegraphics[scale=.5]{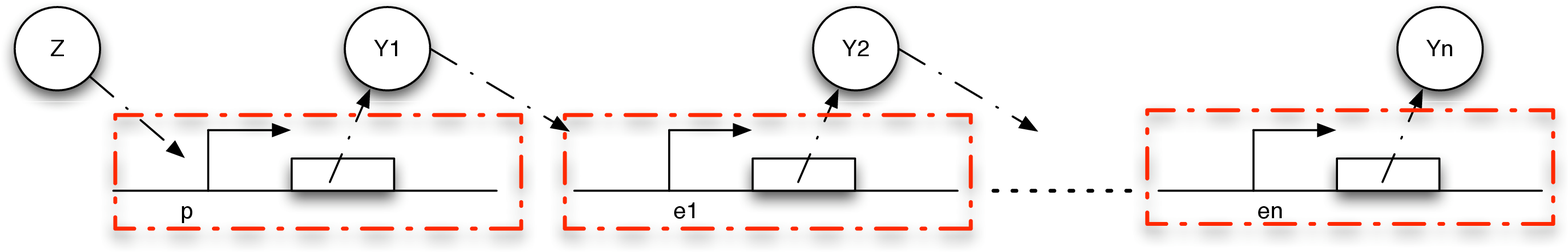}
  \caption{Transcriptional cascade discussed in the text. Each red box contains the transcriptional module described by (\ref{model})}
  \label{fig:cascades}
\end{figure}

A cascade of (infinitesimally) contracting systems is also (infinitesimally)
contracting (see Appendix~\ref{sec:appendix_cascades} for the proof).
This implies that any transcriptional cascade, will also give rise to a
contracting system, and, in particular, will entrain to periodic inputs.
By a transcriptional cascade we mean a system as shown in 
Figure~\ref{fig:cascades}.
In this figure, we interpret the intermediate variables $Y_i$ as transcription
factors, making the simplifying assumption that TF concentration is proportional
to active promoter for the corresponding gene.  (More complex models,
incorporating transcription, translation, and post-translational modifications
could themselves, in turn, be modeled as cascades of contracting systems.)

\subsubsection{More abstract systems}

We can extend our results even further, to a larger class of nonlinear systems,
as long as the same general structure is present.  This can be useful for example to design new synthetic transcription modules or to analyze the entrainment properties of general biological systems. We start with a discussion
of a two dimensional system of the form:
\begin{equation} \label{general_model}
\begin{array}{*{20}l}
\dot x \;=\; u \left(t\right)-a \left( x \right) +f \left(x,y\right),\\
\dot y\;=\; -f \left(x,y\right).\\
\end{array}.
\end{equation}
In molecular biology, $a(x)$ would typically represent a
nonlinear degradation, for instance in Michaelis-Menten form, while the
function $f$ represents the interaction between $x$ and $y$.
The aim of this Section is to find conditions on the degradation and
interaction terms that allow one to show contractivity of the unforced
(no input $u$) system, and hence existence of globally attracting limit cycles.

We assume that the state space $C$ is compact (closed and bounded) as well as
convex.

\begin{Theorem} \label{general_thm_1}
System~(\ref{general_model}), without inputs $u$, evolving on a convex compact subset of phase space is contracting, provided
that the following conditions are all satisfied, for each $x,y \in C$:
\begin{itemize}
\item $\frac{\partial a}{ \partial x} >0$;
\item $\frac{\partial f}{ \partial y} >0$;
\item $\frac{\partial f}{ \partial x} $ does not change sign;
\item $\frac{\partial a}{ \partial x}  > 2 \frac{\partial f}{\partial x}$.
\end{itemize}
\end{Theorem}
\noindent
Notice that the last condition is automatically satisfied if
$\frac{\partial f}{ \partial x} <0$, because
$\frac{\partial a}{ \partial x} >0$.

\proof
As before, we prove contraction by constructing an appropriate negative
measure for the Jacobian of the vector field.
In this case, the Jacobian matrix is:
\begin{equation} \label{Jac_general}
J=\left[ \begin{array}{*{20}c}
-\frac{\partial a}{\partial x}+\frac{\partial f}{\partial x} & \frac{\partial f}{\partial y} \\
-\frac{\partial f}{\partial x} & -\frac{\partial f}{\partial y}\\
\end{array} \right].
\end{equation}
Once again, as matrix measure we will use:
\begin{equation} \label{general_lab}
\mu_{P,1}\left(J\right)=\mu_1 \left( PJP^{-1}\right),
\end{equation}
with
\begin{equation}
P=\left[ \begin{array}{*{20}c}
p_1 & 0\\
0 & p_2\\
\end{array}\right],
\end{equation}
and $p_1,p_2 >0$ appropriately chosen.

Using (\ref{general_lab}) we have
\begin{equation}\label{general_meas}
\mu_{P,1} \left(J\right)=\max 
\left\{-\frac{\partial a}{\partial x}+ \frac{\partial f}{\partial x}+  
\abs{\frac{p_2}{p_1}\frac{\partial f}{ \partial x}} \,;\, 
-
\frac{\partial f}{\partial y}+
\abs{\frac{p_1}{p_2}\frac{\partial f}{\partial y}} \right\}.
\end{equation}
Following the same steps as the proof of Theorem~\ref{theorem:contract}, we have to show that:
\begin{equation} \label{ineq_general_1}
 - \frac{\partial f}{\partial y}+ \abs{ \frac{p_1}{p_2}\frac{\partial f}{\partial y}} < -c_1^2,
\end{equation}
\begin{equation} \label{ineq_general}
-\frac{\partial a}{\partial x}+ \frac{\partial f}{\partial x}+  \abs{ \frac{p_2}{p_1}\frac{\partial f}{ \partial x}} < -c_2^2.
\end{equation}

Clearly, if $\partial f/ \partial y >0$ for every $x,y \in C$ and
$p_1<p_2$, the first inequality is satisfied, with 
$$
c_1^2 = \left(\frac{p_1}{p_2}-1\right)\frac{\partial f}{\partial x}.
$$

To prove the theorem we need to show that there exists $p_1<p_2$ and $c_2^2$ satisfying (\ref{ineq_general}).  For such inequality, since $\partial f / \partial x$ does not change sign in $C$ by hypothesis, we have two possibilities:
\begin{enumerate}
\item $\frac{\partial{f}}{\partial x}<0$, $\forall x,y \in C$;
\item $\frac{\partial{f}}{\partial x}>0$, $\forall x,y \in C$.
\end{enumerate}
In the first case, the right hand side of (\ref{ineq_general}) becomes
\begin{equation} \label{ineq_general_2}
-\frac{\partial a}{\partial x}+ \frac{\partial f}{\partial x}- \frac{p_2}{p_1}\frac{\partial f}{ \partial x}
\end{equation}
Choosing $p_2/p_1=1+
\varepsilon$, with $\varepsilon >0$, we have:
$$
-\frac{\partial a}{\partial x}+ \frac{\partial f}{\partial x}- \frac{p_2}{p_1}\frac{\partial f}{ \partial x} = -\frac{\partial a}{\partial x} + \varepsilon \frac{\partial f}{ \partial x}.
$$

Specifically, if we now pick
\[
\varepsilon > \frac{A}{B}
\]
where
$A = \max \frac{\partial a}{\partial x}$
and
$B = \min\abs{\frac{\partial f}{\partial x}}$, we have that the above quantity is uniformly negative definite, i.e.
$$
\exists c_{2,1}^2 : \quad -\frac{\partial a}{\partial x} + \varepsilon \frac{\partial f}{ \partial x}<-c_{1,2}^2.
$$

In the second case, the right hand side of (\ref{ineq_general}) becomes
\begin{equation} \label{ineq_general_3}
-\frac{\partial a}{\partial x}+ \frac{\partial f}{\partial x}+
\frac{p_2}{p_1}\frac{\partial f}{ \partial x}.
\end{equation}
Again, by choosing  $p_2/p_1=1+ \varepsilon$, with $\varepsilon >0$, we have
the following upper bound for the expression in~(\ref{ineq_general_3}):
\begin{equation} \label{ineq_general_3b}
-\frac{\partial a}{\partial x}+ 2\frac{\partial f}{\partial x}+ \varepsilon
\frac{\partial f}{ \partial x}.
\end{equation}
Thus, it follows that $\mu_{P,1}\left(J\right)<-c^2$
provided that the above quantity is uniformly negative definite. Since, by hypotheses,
\begin{equation} \label{final}
\frac{\partial a}{\partial x} > 2 \frac{\partial f}{\partial x} \quad \forall x,y \in C,
\end{equation}
then  $\exists c_{2,2}^2 : \quad -\frac{\partial a}{\partial x}+ \frac{\partial f}{\partial x}+
\frac{p_2}{p_1}\frac{\partial f}{ \partial x}\le -c_{2,2}^2$.
The proof of the Theorem is now complete.
\endproof

From a biological viewpoint, the hardest
  hypothesis to satisfy in Theorem
\ref{general_thm_1} might be that on the derivatives of
$f\left(x,y\right)$. However, it is possible to relax the hypothesis on
$\partial f / \partial x$ if the rate of change of $a \left(x\right)$ with
respect to $x$, i.e. $\partial a / \partial x$, is sufficiently larger than
$\partial f / \partial x$. In particular, the following result can be proved.
\begin{Theorem} \label{general_thm_2}
System~(\ref{general_model}), without inputs $u$, evolving on a convex compact set, is contractive
provided that:
\begin{itemize}
\item $\partial a/ \partial x >0$, $\forall x \in C$;
\item $\partial f/ \partial y >0$, $\forall x,y \in C$;
\item $\partial a / \partial x > \max_{C} \left\{ 2 \abs{ \partial f/ \partial x } \right\}\ $.
\end{itemize}
\end{Theorem} 
\proof
The proof is similar to that of Theorem \ref{general_thm_1}. In particular, we can repeat the same derivation to obtain again inequality (\ref{ineq_general}). Thence, as no hypothesis is made on the sign of 
$\partial f / \partial x$,
choosing $p_2/p_1 =1+\varepsilon$ we have
\begin{equation} \label{ineq_general_thm_n}
-\frac{\partial a}{\partial x}+ \frac{\partial f}{\partial x}+  
\abs{\frac{p_2}{p_1}\frac{\partial f}{ \partial x}} 
= -\frac{\partial a}{\partial x}+ \frac{\partial f}{\partial x}+  \abs{ \frac{\partial f}{ \partial x} } + \varepsilon \abs{ \frac{\partial f}{ \partial x} }.
\end{equation}
Thus, it follows that, if $\partial a / \partial x \ge 2 \abs{ \partial f / \partial x}$,
then $\exists$ $ c^2$ such that $\mu_{P,1}\left(J\right)<-c^2$, implying contractivity. The above
condition is satisfied by hypotheses, hence the theorem is proved.
\endproof

\subsubsection*{Remarks} \label{fur_exten}

Theorems~\ref{general_thm_1} and~\ref{general_thm_2} show the
possibility of designing with high flexibility the self-degradation and
interaction functions for an input-output module. 

This flexibility can be further increased, for example in the following ways:

\begin{itemize}
\item Results similar to that of the above Theorems can be derived (and also
  extended) if some self degradation rate for $y$ is present in
  (\ref{general_model}), i.e.
\begin{equation}
\begin{array}{*{20}l}
\dot x = u \left(t\right)-a \left( x \right) +f \left(x,y\right)\\
\dot y= -b\left(y\right)-f \left(x,y\right)\\
\end{array}
\end{equation}
with $\frac{\partial b}{\partial y}<0$.
\item Theorem \ref{general_thm_1} and Theorem \ref{general_thm_2} can also be
  extended to the case in which the $X$-module drives more than one downstream
  transcriptional modules.
\end{itemize}

\section{Materials and Methods}

All simulations are performed in MATLAB (Simulink), Version 7.4, with variable step ODE solver ODE23t. Simulink models are available upon request.

\section{Conclusions}

We have presented a systematic methodology to derive conditions for
transcriptional modules to be globally entrained to periodic inputs. By means
of contraction theory, a useful tool from dynamical systems, we showed that it
is possible to use non-Euclidean norms and their associated matrix measures to
characterize the behavior of several modules when subject to external periodic
excitations. Specifically, starting with a simple bimolecular reaction, we
considered the case of a general externally-driven transcriptional module and
extended the analysis to some important generalizations including the case of
multiple driven systems. In all cases conditions are derived by proving that
the module of interest is contracting under some generic assumptions on its
parameters. The importance of the results presented in the paper from a design
viewpoint are also discussed by means of more abstract systems where generic
nonlinear degradation and interaction terms are assumed.

\appendix

\section{$K$-reachable sets}

We will make use of the following definition:
\begin{Definition}\label{defn:K-reachable}
Let $K>0$ be any positive real number.
A subset $C\subset\R^n$ is \emph{$K$-reachable} if, for any two points $x_0$ and
$y_0$ in $C$ there is some continuously differentiable curve 
$\gamma : \left[  0, 1 \right] \rightarrow C$ 
such that:
\begin{enumerate}
\item
$\gamma \left(0\right) = x_0$, 
\item
$\gamma\left(1\right)=y_0$ and 
\item
$\abs{\gamma '  \left(r\right)} \le K \abs{y_0 - x_0}$, $\forall r$. 
\end{enumerate}
\end{Definition}

For convex sets $C$, we may pick $\gamma(r)=x_0+r(y_0-x_0)$, so 
$\gamma'(r)= y_0-x_0$ and we can take $K=1$.  Thus, convex sets are
$1$-reachable, and it is easy to show that the converse holds as well.

Notice that a set $C$ is $K$-reachable for some $K$ if and only if the length
of the geodesic (smooth) path (parametrized by arc length), connecting any two
points $x$ and $y$ in $C$, is bounded by some multiple $K_0$ of the Euclidean
norm, $\abs{y-x}_2$. Indeed, re-parametrizing to a path $\gamma$ defined on
$\left[0,1\right]$, we have: 
\[
\abs{\gamma ' \left(r\right)}_2 \le K_0 \abs{y-x}_2.
\]
Since in finite dimensional spaces all the norms are equivalent, then it is possible to obtain a suitable $K$ for Definition \ref{defn:K-reachable}.

\begin{Remark}
The notion of $K$-reachable set is weaker that that of convex set. 
Nonetheless, in Theorem \ref{theo:k-reachable}, we will prove that
trajectories of a smooth system, evolving on a $K$-reachable set,
converge towards each other, even if $C$ is not convex.
This additional generality allows one to establish contracting behavior for
systems evolving on phase spaces exhibiting ``obstacles'', as are frequently
encountered in path-planing problems, for example.  
A mathematical example of a set with obstacles follows.
\end{Remark}

\begin{Example}
Consider the two dimensional set, $C$, defined by the following contraints:
$$
\begin{array}{*{20}c}
x^2+y^2 \ge 1, & x \ge 0, & y \ge 0 \\
\end{array}.
$$
Clearly, $C$ is a non-convex subset of $\R^2$.
We claim that $C$ is $K$-reachable, for any positive real number
$K>\frac{2}{\pi}$.
Indeed, given any two points $a$ and $b$ in $C$, there are two possibilities:
either the segment connecting $a$ and $b$ is in $C$, or it intersects the unit
circle.  In the first case, we can simply pick the segment as a curve
($K=1$).  In the second case, one can consider a straight segment that is
modified by taking the shortest perimiter route around the circle; the length
of the perimeter path is at most $\frac{2}{\pi}$ times the length of the
omitted segment.  (In order to obtain a differentiable, instead of merely a
piecewise-differentiable, path, an arbitrarily small increase in $K$ is needed.)
\end{Example}

\section{Proof of Theorem \ref{theo:main}}

We now prove the main result on contracting systems, i.e. Theorem \ref{theo:main}, under the hypotheses that the set $C$, i.e. the set on which the system evolves, is $K$-reachable.

\begin{Theorem}\label{theo:k-reachable}
Suppose that $C$ is a $K$-reachable subset of $\R^n$ and that $f(t,x)$ is
infinitesimally contracting with contraction rate $c^2$.
Then, for every two solutions $x(t)=\varphi(t,0,\xi )$ and $z(t)=\varphi(t,0,\zeta )$
it holds that:
\be{eqn:weak}
\abs{x(t)-z(t)} \leq  K e^{-c^2t} \abs{\xi -\zeta } 
\quad\quad\forall\,t\ge 0\,.
\ee
\end{Theorem}
\proof
Given any two points $x\left(0\right)= \xi$ and $z\left(0\right)= \zeta$ in
$C$, pick a smooth curve $\gamma : \left[ 0, 1\right] \rightarrow C$, such
that $\gamma \left(0\right)=\xi$ and $\gamma \left(1\right)=\zeta$. Let  
$\psi\left(t,r\right)=\varphi(t,0,\gamma \left(r\right)$, that is,
the solution of system (\ref{eqn:gensys}) rooted in 
$\psi\left(0,r\right) =\gamma \left(r\right)$, $r \in [0,1]$. 
Since $\varphi$ and $\gamma$ are continuously differentiable, also
$\psi\left(t,r\right)$ is continuously differentiable in both arguments.
We define
\[
w(t,r) := \frac{\partial \psi}{\partial r}(t,r).
\]
It follows that
$$
\frac{\partial w}{\partial t}(t,r) = \frac{\partial}{\partial t}\left(\frac{\partial \psi}{\partial r}\right)=\frac{\partial}{\partial r}\left(\frac{\partial \psi}{\partial t}\right)=\frac{\partial}{\partial r}f(\psi \left(t,r\right),t).
$$
Now,
$$
\frac{\partial}{\partial r}f(\psi \left(t,r\right),t) = \frac{\partial f}{\partial x}(\psi\left(t,r\right),t)\frac{\partial \psi}{\partial r}(t,r)
$$
so, we have:
\begin{equation}
\frac{\partial w}{\partial t}(t,r) = J(\psi\left(t,r\right),t) w(t,r),
\end{equation}
where $J(\psi\left(t,r\right),t) = \frac{\partial f}{\partial x}(\psi\left(t,r\right),t)$.
Using  Coppel's inequality \cite{Vid_93}, yields
\begin{equation} \label{eq:coppelthm}
\abs{w(t,r)} \le \abs{w(0,r)}  e^{\int_{0}^t \mu \left( J\left(\tau\right)\right)d\tau}\leq K \abs{\xi-\zeta} e^{-c^2 t},
\end{equation}
$\forall x \in C$, $\forall t \in \R^+$, and $\forall r \in [0,1]$. 
Notice the Fundamental Theorem of Calculus,
we can write
$$
\psi\left(t,1\right)-\psi\left(t,0\right) = \int_0^1{w(t,s)} ds .
$$
Hence, we obtain
$$
\abs{x(t)-z(t)} \leq \int_0^1{\abs{w(t,s)}}ds.
$$
Now, using (\ref{eq:coppelthm}), the above inequality becomes:
$$
\abs{x(t)-z(t)} \le \int_0^1 \left(\abs{w(0,s)}  e^{\int_{0}^t \mu \left( J\left(\tau\right)\right)d\tau} \right) ds \le K \abs{\xi -\zeta}e^{-c^2t}.
$$
The Theorem is then proved.
\endproof

\noindent{\bf Proof of Theorem~\ref{theo:main}:}
The proof follows trivially from Theorem \ref{theo:k-reachable}, after having noticed that in the convex case, we may assume $K=1$.\qed

\section{Proof of Theorem \ref{theorem:contraction}}
In this Section we assume that the vector field $f$ is $T$-periodic and prove Theorem \ref{theorem:contraction}.

Before starting with the proof of Theorem \ref{theorem:contraction} we make the following:
\begin{Remark}\label{rem:periodicf}
Periodicity implies that the initial
time is only relevant modulo $T$.  More precisely:
\be{eqn:periodicf}
\varphi(kT+t,kT,\xi ) = \varphi(t,0,\xi )
\quad\quad\forall\,k\in \N,t\geq 0,\,x\in C\,.
\ee
Indeed, let $z(s)=\varphi(s,kT,\xi )$, $s\geq kT$, and consider the function
$x(t)=z(kT+t)=\varphi(kT+t,kT,\xi )$, for $t\geq 0$.
So,
\[
\dot x(t) = \dot z(kT+t) = f(kT+t,z(kT+t)) = f(kT+t,x(t)) = f(t,x(t))\,,
\]
where the last equality follows by $T$-periodicity of $f$.
Since $x(0)=z(kT)=\varphi(kT,kT,\xi )=\xi $, it follows by uniqueness of
solutions that $x(t)=\varphi(t,0,\xi ) = \varphi \left(kT+t,kT,\xi\right)$, which is~(\ref{eqn:periodicf}).
As a corollary, we also have that
\be{eqn:periodicfc}
\varphi(kT+t,0,\xi ) = \varphi(kT+t,kT,\varphi(kT,0,\xi )) = \varphi(t,0,\varphi(kT,0,\xi ))
\quad\quad\forall\,k\in \N,t\geq 0,\,x\in C
\ee
where the first equality follows from the semigroup property of solutions (see e.g. \cite{mct}), and
the second one from~(\ref{eqn:periodicf}) applied to
$\varphi(kT,0,\xi )$ instead of $\xi $.
\end{Remark}

Define now
\[
P(\xi ) = \varphi(T,0,\xi ),
\]
where $\xi = x \left(0\right) \in C$. The following Lemma will be useful in what follows.

\begin{Lemma}\label{lemma:P}
$P^k(\xi ) = \varphi(kT,0,\xi )$ for all $k\in \N$ and $\xi \in C$.
\end{Lemma}
\proof
We will prove the Lemma by recursion. In particular, the statement is true by definition when $k=1$.
Inductively, assuming it true for $k$, we have:
\[
P^{k+1}(\xi ) = 
P(P^k(\xi )) =
\varphi(T,0,P^k(\xi )) = 
\varphi(T,0,\varphi(kT,0,\xi )) =
\varphi(kT+T,0,\xi )\,,
\]
as wanted.
\endproof

\begin{Theorem}\label{theo:contraction-general}
Suppose that:
\begin{itemize}
\item $C$ is a closed $K$-reachable subset of $\R^n$;
\item $f$ is infinitesimally contracting with contraction rate $c^2$;
\item $f$ is $T$-periodic;
\item $Ke^{-c^2T}<1$.
\end{itemize}
Then, there is an unique periodic solution $\alpha (t):[0,\infty )\rightarrow C$ of~(\ref{eqn:gensys}) having period $T$. Furthermore, every solution $x(t)$, such that $x\left(0\right)=\xi \in C$, converges to $\alpha \left(t\right)$, i.e. $\abs{x(t)-\alpha (t)}\rightarrow 0$ as $t\rightarrow \infty $.
\end{Theorem}
\proof
Observe that $P$ is a contraction with factor $Ke^{-c^2T}<1$: 
$\abs{P(\xi )-P(\zeta )}\leq Ke^{-c^2T} \abs{\xi-\zeta}$
for all $\xi ,\zeta \in C$,
as a consequence of Theorem~\ref{theo:k-reachable}.
The set $C$ is a closed subset of $\R^n$ and hence complete as a metric space
with respect to the distance induced by the norm being considered.
Thus, by the contraction mapping theorem, there is a (unique) fixed
point $\bar \xi $ of $P$.
Let $\alpha (t):=\varphi(t,0,\bar \xi )$.
Since $\alpha (T)=P(\bar \xi )={\bar \xi }=\alpha (0)$, $\alpha (t)$ is a periodic orbit of
period $T$. 
Moreover, again by Theorem~\ref{theo:k-reachable},
we have that
$\abs{x(t)-\alpha (t)} \leq  K e^{-c^2t} \abs{\xi -\bar \xi }\rightarrow 0$.
Uniqueness is clear, since two different periodic orbits would be disjoint
compact subsets, and hence at positive distance from each other,
contradicting convergence. This completes the proof.
\endproof

\noindent{\bf Proof of Theorem~\ref{theorem:contraction}:}
It will suffice to note that the assumption $Ke^{-c^2T}<1$ in Theorem \ref{theo:contraction-general} is automatically satisfied when the set $C$ is convex (i.e. $K=1$) and the system is infinitesimally contracting.\qed

Notice that, even in the non-convex case, the assumption $Ke^{-c^2T}<1$ can be
ignored, if we are willing to assert only the existence (and global
convergence to) a unique periodic orbit, with some period $kT$ for some integer
$k>1$.  Indeed, the vector field is also $kT$-periodic for any integer $k$.
Picking $k$ large enough so that $Ke^{-c^2kT}<1$, we have the conclusion that
such an orbit exists, applying Theorem~\ref{theo:contraction-general}.

\section{Cascades}
\label{sec:appendix_cascades}

In order to show that cascades of contracting systems remain contracting, it
is enough to show this, inductively, for a cascade of two systems.

Consider a system of the following form:
\beqn
\dot  x &=& f(t,x)\\
\dot  y &=& g(t,x,y)
\eeqn
where $x(t)\in C_1\subseteq \R^{n_1}$ and $y(t)\in C_2\subseteq \R^{n_2}$ for all $t$ ($C_1$ and $C_2$ are two $K$-reachable sets).
We write the Jacobian of $f$ with respect to $x$ as
$A(t,x) = \frac{\partial f}{\partial x}(t,x)$,
the Jacobian of $g$ with respect to $x$ as
$B(t,x,y) = \frac{\partial g}{\partial x}(t,x,y)$,
and the Jacobian of $g$ with respect to $y$ as
$C(t,x,y) = \frac{\partial g}{\partial y}(t,x,y)$,

We assume the following:
\begin{enumerate}
\item
The system $\dot x=f(t,x)$ is infinitesimally contracting with respect to some norm
(generally indicated as $\abs{\bullet}_\ast$), with some contraction rate $c_1^2$, that is,
$\mu _\ast(A(t,x))\leq  -c_1^2$
for all $x\in C_1$ and all $t\geq 0$, where $\mu _\ast$ is the matrix measure associated
to $\abs{\bullet}_\ast$.
\item
The system $\dot y=f(t,x,y)$ is infinitesimally contracting with respect to some 
norm (which is, in general different from $\abs{\bullet}_\ast$, and is denoted by $\abs{\bullet}_{\ast \ast}$), with contraction rate $c_2^2$, when $x$ is viewed
a a parameter in the second system, that is,
$\mu _{\ast \ast}(C(t,x,y))\leq  -c_2^2$
for all $x\in C_1$, $y\in C_2$ and all $t\geq 0$, where $\mu _{\ast \ast}$ is the matrix measure
associated to $\abs{\bullet}_{\ast \ast}$.
\item
The mixed Jacobian $B(t,x,y)$ is bounded:
$\norm{B(t,x,y)}\leq k^2$, for all $x\in C_1$, $y\in C_2$ and all $t\geq 0$,
for some real number $k$, where ``$\norm{\bullet}$'' is the
operator norm induced by $\abs{\bullet}_{\ast}$ and $\abs{\bullet}_{\ast \ast}$ on linear
operators $\R^{n_1 \times n_2 \times 1} \rightarrow  \R^{n_1\times n_2}$.
(All norms in Euclidean space being equivalent, this can be verified in any
norm.)
\end{enumerate}

We claim that, under these assumptions, the complete system is infinitesimally
contracting.  More precisely, pick any two positive numbers $p_1$ and $p_2$
such that 
\[
c_1^2 \,-\, \frac{p_2}{p_1} k^2 \;>\; 0
\]
and let
\[
c^2 := \min \left\{c_1^2 - \frac{p_2}{p_1} k^2, c_2^2\right\} \,.
\]
We will show that $\mu (J)\leq  -c^2$, where $J$ is the full Jacobian:
\[
J = 
\left[ 
\begin{array}{*{20}c}
A & 0 \\
B & C
\end{array} \right]
\]
with respect to the matrix measure $\mu $ induced by the following norm in
$\R^{n_1\times n_2}$:
\[
\abs{(x_1,x_2)} = p_1 \abs{x_1}_{\ast} + p_2 \abs{x_2}_{\ast \ast} \,.
\]
Since
\[
(I+hJ)x = 
\left[ 
\begin{array}{*{20}c}
(I+hA)x_1  \\
hBx_1 + (I+hC)x_2
\end{array} \right]
\]
for all $h$ and $x$, we have that, for all $h$ and $x$:
\beqn
\abs{(I+hJ)x}
&=&
p_1\abs{(I+hA)x_1} + p_2 \abs{hBx_1 + (I+hC)x_2}\\
&\leq &
p_1\abs{I+hA}\abs{x_1} +
p_2\abs{hB}\abs{x_1} +
p_2\abs{I+hC}\abs{x_2},
\eeqn
where from now on we drop subscripts for norms.
Pick now any $h>0$
and a unit vector $x$ (which depends on $h$)
such that $\norm{I+hJ} = \abs{(I+hJ)x}$.
Such a vector $x$ exists by the definition of induced matrix norm, and we note
that $1= \abs{x} = p_1\abs{x_1}_{\ast}+p_2\abs{x_2}_{\ast \ast}$, by the definition of the norm
in the product space.
Therefore:
\beqn
\frac{1}{h} \left(\norm{I+hJ}-1\right)
&=&
\frac{1}{h} \left(\abs{(I+hJ)x}-\abs{x}\right)\\
&\leq &
\frac{1}{h} 
\left(
p_1\abs{I+hA}\abs{x_1} +
p_2\abs{hB}\abs{x_1} +
p_2\abs{I+hC}\abs{x_2}
- p_1 \abs{x_1} - p_2 \abs{x_2}
\right)\\
&=&
\frac{1}{h} 
\left(\abs{I+hA} -1 + \frac{p_2}{p_1}h\abs{B} \right) p_1\abs{x_1}
+
\frac{1}{h}
\left(\abs{I+hC} - 1 \right) p_2\abs{x_2}\\
&\leq &
\max\left\{
\frac{1}{h}\left(\abs{I+hA} -1\right) + \frac{p_2}{p_1}k^2
\,,\,
\frac{1}{h}\left(\abs{I+hC} - 1 \right)
\right\} \,,
\eeqn
where the last inequality is a consequence of the fact that
$\lambda _1a_1+\lambda _2a_2\leq \max\{a_1,a_2\}$ for any nonnegative numbers
with $\lambda _1+\lambda _2=1$ (convex combination of the $a_i$'s).
Now taking limits as $h\searrow0$, we conclude that
\[
\mu (J)\leq 
\max \left\{-c_1^2 + \frac{p_2}{p_1} k^2, -c_2^2\right\}
=
-c^2 \,,
\]
as desired.

\section{A counterexample to entrainment}

In~\cite{eds:arxiv09} there is given an example of a system with the following
property: when the external signal $u(t)$ is constant, all solutions converge
to a steady state; however, when $u(t)=\sin t$, solutions become chaotic.
(Obviously, this system is not contracting.)
The equations are as follows:
\def\lorx{\xi }
\def\lory{\psi }
\def\lorz{\zeta }
\beqn
\dot x&=&- x -u\\
\dot p&=&-p + \alpha (x+u)\\
\dot \lorx &=& 10(\lory-\lorx)\\
\dot \lory &=& 28p\lorx- \lory- p \lorx \lorz\\
\dot \lorz &=& p \lorx \lory - (8/3) \lorz
\eeqn
where $\alpha (y)=y^2/(K+y^2)$ and $K=0.0001$.
Figure~\ref{fig:example2} shows typical solutions of this system with a
periodic and constant input respectively.
The function ``rand'' was used in MATLAB to produce random values in the
range $[-10,10]$.
\begin{figure}[h,t]
\begin{center}
\includegraphics[scale=0.5]{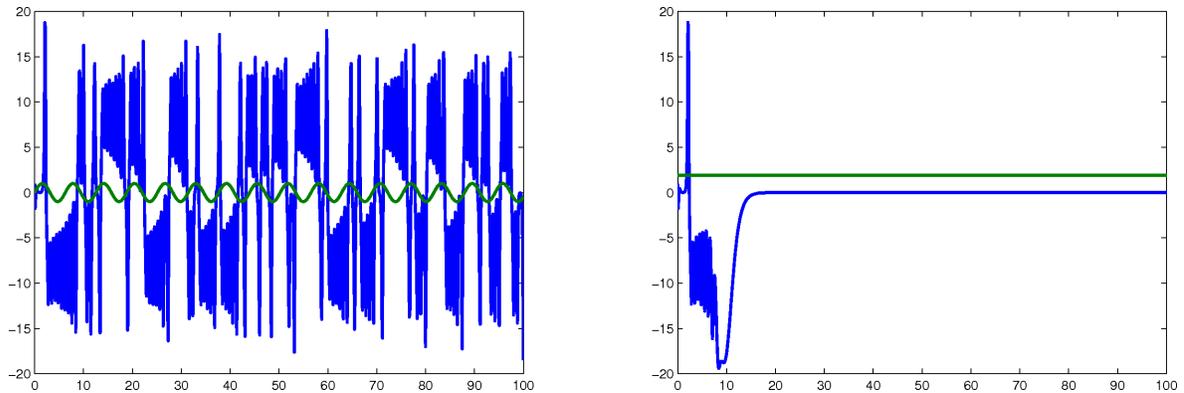}
\caption{Simulation of counter-example,
done with the following randomly-chosen input and initial conditions:
$u(t)=1.89$, 
$x(0)=2.95$ $p(0)=-0.98$, $\lorx(0)=0.94$, $\lory(0)=-4.07$, $\lorz(0)=4.89$.
Green: inputs are $u(t)=\sin t$ (left panel) and $u(t)=5.13$ (randomly picked,
right panel).
Blue: $\lorx(t)$.
Note chaotic-like behavior in response to periodic input, but steady state
in response to constant input.}
\label{fig:example2}
\end{center}
\end{figure}

\end{document}